\documentclass{PoS}
\usepackage{graphicx, color, subfigure, slashed, listings, fancyhdr, amsmath, amsthm, amssymb, bbm}

\title{The Quark-Gluon Vertex in Landau gauge QCD}

\ShortTitle{The Quark-Gluon Vertex in Landau gauge QCD}

\author{\speaker{Markus Hopfer}\\
        Karl-Franzens-University Graz\\
        E-mail: \email{markus.hopfer@uni-graz.at}}

\author{Andreas Windisch\\
        Karl-Franzens-University Graz\\
       E-mail: \email{andreas.windisch@uni-graz.at}}

\author{Reinhard Alkofer\\
        Karl-Franzens-University Graz\\
       E-mail: \email{reinhard.alkofer@uni-graz.at}}

\abstract{The coupled system of the quark-gluon vertex and quark propagator 
Dyson-Schwinger equations (DSEs) is investigated within Landau gauge QCD. 
The aim is to get a deeper insight into the mechanisms of quark confinement 
and dynamical chiral symmetry breaking and into a possible relation 
between these two phenomena.
To this end an earlier study is extended by improving systematically on the 
truncations imposed on the quark-gluon vertex DSE. A clear infrared enhancement 
for all tensor structures of the  quark-gluon vertex is obtained.}

\FullConference{Xth Quark Confinement and the Hadron Spectrum,\\
		October 8-12, 2012\\
		TUM Campus Garching, Munich, Germany}

\begin{document}

%%%%%%%%%%%%%%%%%%%%%%%%%%%%%%%%%%%%%%%%%%%%%%%%%%%%%%%%%%%%%%%%%%%%%%%%%%%%%%%%%%%%%%%%%%%%%%%%%%%%%%  

\section{Motivation}
\label{sec:intro}
The mechanisms behind the most prominent features of QCD,
namely confinement and dynamical chiral symmetry breaking 
(D$\chi$SB), are still not completely understood. 
Moreover, a possible relation between these two phenomena is by no means obvious. Nevertheless, 
lattice simulations \cite{Karsch:2003jg,Aoki:2006br} performed at finite temperature show that the
corresponding phase transitions lie remarkably close together, which is a surprising result
since the confinement phase transition, indicated by a breaking of center symmetry, is
mainly driven by gluodynamics as pointed out in \cite{Braun:2009gm},
whereas strong quark interactions seem to be  
responsible for the chiral phase transition. Moreover, a certain 
quark-gluon interaction strength is needed 
in order to generate mass dynamically in the infrared region.
The aim is to shed more light on these effects 
by studying the quark-gluon vertex since this object is the link between the Yang-Mills and the matter sector
of the theory.
As both phenomena are related to the low-energy region of the theory a non-perturbative framework 
is necessary in addition to perturbative techniques. Dyson-Schwinger equations (DSEs) are an appropriate 
tool to explore the theory over a wide scope of momenta ranging from the deep infrared up to the perturbative
regime, see {\it e.g.}, ref.\ \cite{Alkofer:2000wg,Maris:2003vk} and references therein.
They build up to an infinite tower of coupled 
integral equations such that carefully chosen truncations have to be applied in order
to evaluate the equations. Therefore, at some point a cooperation 
with other non-perturbative methods
such as lattice simulations is necessary
 in order to render the Dyson-Schwinger framework reliable and robust and to acquire results
which might be too difficult to obtain from a lattice calculation.

In these proceedings we extend previous studies of the
quark-gluon vertex \cite{Alkofer:2008tt,Windisch:2012de} by systematically 
improving on the employed truncations. The starting point is the coupled system of the quark propagator
and quark-gluon vertex DSE at vanishing temperatures 
in the Landau gauge using as input for the gluon propagator 
either fits \cite{Alkofer:2008tt,Fischer:2002hna,Fischer:2003rp} or results from a self-consistent 
treatment of the corresponding system of DSEs 
\cite{Hopfer:2012ht}.\footnote{These fits correspond to a 
so-called scaling solution, one of two solution types 
of Landau gauge Green functions when classified by their infrared properties, 
see {\it e.g.}, ref.\ 
\cite{Fischer:2008uz} and references therein. However, we want to stress 
the evidence that the differences of these types of solutions 
in the far infrared are irrelevant for phenomenological results.
\newline
Furthermore, using numerical input obtained from a self-consistent solution of the 
corresponding Yang-Mills system
leaves additional freedom in taking novel findings from other investigations, 
see {\it e.g.}, ref. \cite{Huber:2012kd}, into account.} 
The three-gluon vertex model used in \cite{Alkofer:2008tt,Windisch:2012de} has been modified in order 
to take recent lattice results \cite{Cucchieri:2008qm} into account. 

Although the ghost propagator does not enter directly the DSEs for the quark propagator and the
quark-gluon vertex one has to note that 
the infrared suppression of the gluon propagator is a direct consequence of the  
the infrared enhancement of the ghost propagator (see {\it e.g.}, ref.\ \cite{Watson:2001yv}). 
The resulting positivity  violation for transverse gluons identifies 
the role of the transverse gluons in a  non-perturbative 
BRST quartet mechanism \cite{Alkofer:2011pe} and thus a partial aspect of gluon 
confinement: In Landau gauge QCD transverse gluons are confined and therefore not confining.
Consequently,  due to the infrared suppression of the gluon propagator the quark-gluon vertex has to
acquire a certain strength in the infrared to provide D$\chi$SB or maybe even the infrared
divergence indicating quark confinement. Within the truncation of ref.\ \cite{Alkofer:2008tt} 
(and at vanishing temperatures and densities) D$\chi$SB
and quark confinement occur either together in a self-consistent way or are both absent.

An improved truncation of the quark-gluon vertex
DSE might give deeper insights into the possible relation between quark confinement and
D$\chi$SB. 
In addition, isolating important tensor structures of the quark-gluon vertex 
will provide a guidance when modeling this object in phenomenological applications.
In particular, corresponding investigations 
at non-vanishing temperatures and/or quark chemical potentials
involving the quark-gluon vertex can benefit from a
better understanding of the vertex structure, see {\it e.g.}, ref.~\cite{Hopfer:2012qr}.
%%%%%%%%%%%%%%%%%%%%%%%%%%%%%%%%%%%%%%%%%%%%%%%%%%%%%%%%%%%%%%%%%%%%%%%%%%%%%%%%%%%%%%%%%%%%%%%%%%%%%%%%%%%%%%%%%
\section{The coupled system of equations}
\label{sec:coupled_system}
\begin{figure}[h]
\centerline{%
\includegraphics[width=9cm]{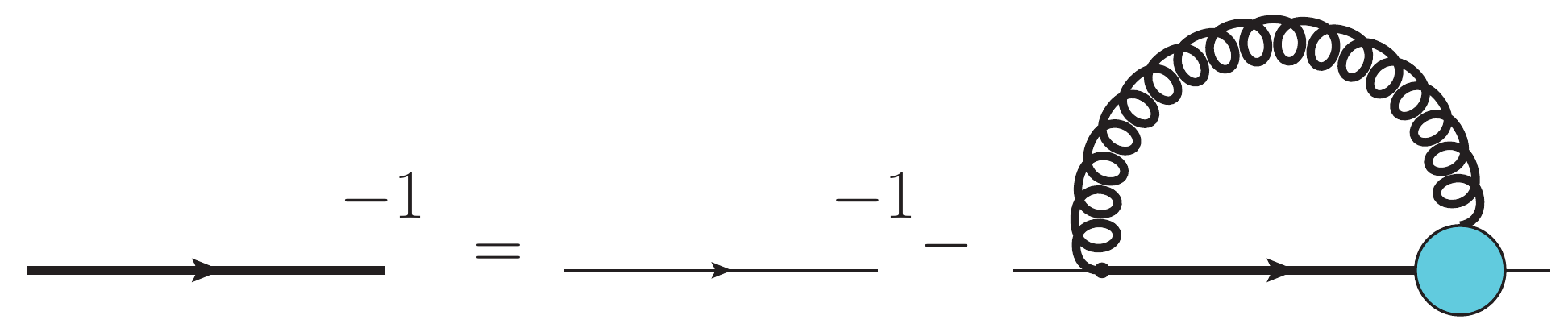}\\[0.5cm]}
\centerline{%
\includegraphics[width=11.5cm]{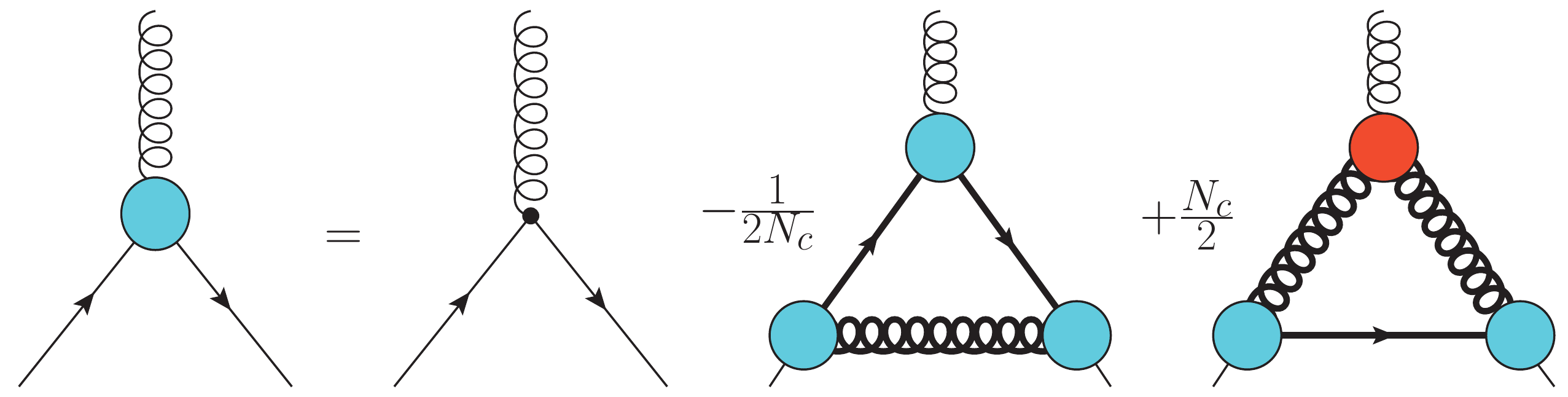}}
\caption{The quark propagator and the quark-gluon vertex DSE.
All internal propagators are dressed. The vertex equation has been derived from a 
3PI effective action \cite{Berges:2004pu}. The color prefactors of the Abelian- and the non-Abelian
diagram are shown explicitely.}
\label{Fig1}
\end{figure}

The coupled system of equations for the quark propagator and the quark-gluon vertex is 
depicted in Fig. \ref{Fig1} 
and serves as a starting point for this investigation.
The equation for the quark-gluon vertex has been derived from a 3PI effective action
and corresponds to the DSE truncated at three-point level \cite{Alkofer:2008tt,Berges:2004pu}. Note that 
all vertices are dressed due to the nPI-based approach. The last two diagrams on the 
right-hand side of the vertex equation are referred to as the \textit{Abelian} and 
the \textit{non-Abelian} diagram. The color traces yield a prefactor of 
${N_c}/{2}$ for the non-Abelian diagram, whereas the Abelian diagram is suppressed by a 
factor of $-{1}/{2N_c}$. This natural $N_c^2$ difference between the two diagrams has been found
to be even more severe due to the dynamics 
of the system \cite{Alkofer:2008tt}.\footnote{Furthermore, one can argue that the two dressed 
quark propagators in the Abelian diagram
act as an additional damping factor not only for very heavy quarks but also in the chiral
limit via $D\chi SB$. Of course, only a detailed investigation of the Abelian diagram in 
upcoming studies will reveal
whether this naive argument also holds in a self-consistent treatment of the system.} 
Indeed, the simplest possible self-consistent treatment of the system 
reveals a suppression of the Abelian diagram by two orders of magnitude compared to the non-Abelian
diagram. In this scenario the quark-gluon vertex DSE has been treated within a 
\textit{leading order skeleton expansion}, i.e. all dressed vertices on the right-hand side 
of the vertex equation are substituted by their bare counterparts. We note however that in
this setting the resulting feed-back on the propagator, i.e. the effective 
quark-gluon interaction strength, is not strong enough to trigger $D\chi SB$. This result on the one
hand confirms previous DSE studies \cite{LlanesEstrada:2004jz,Alkofer:2008tt} and on the other hand makes
the necessity obvious to treat the quark-gluon vertex DSE in a self-consistent manner including
at least the most important tensor structures relevant for $D\chi SB$ in order to generate mass
dynamically. 

Thus, the road map towards a self-consistent solution of the system depicted in Fig. \ref{Fig1} 
contains the following points. In a previous first step only parts of the non-Abelian diagram 
have been taken into account \cite{Windisch:2012de}, whereas 
in these proceedings now the full non-Abelian diagram has been included 
into the calculations. A detailed investigation of the Abelian diagram within a self-consistent
treatment will be performed in an upcoming study.

\subsection{Towards a self-consistent solution}
\label{sec:towards_a_solution}
\begin{figure}[h]
\centerline{%
\includegraphics[width=9cm]{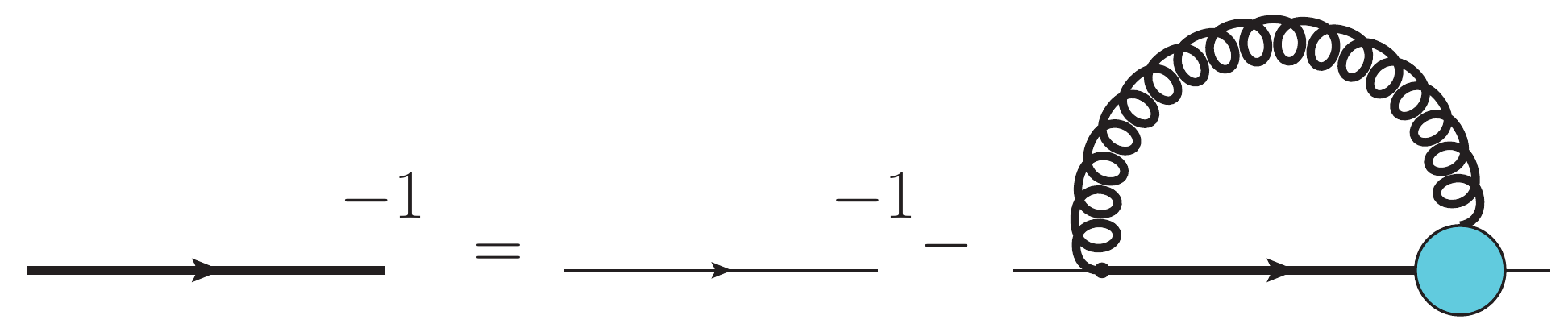}}
\caption{The quark propagator DSE coupled to a full quark-gluon vertex DSE denoted by the blue blob. 
The gluon propagator has been taken as input from external DSE calculations 
\cite{Alkofer:2008tt,Fischer:2002hna,Hopfer:2012ht}.}
\label{Fig2}
\end{figure}

\begin{figure}[h]
\centerline{%
\includegraphics[width=9cm]{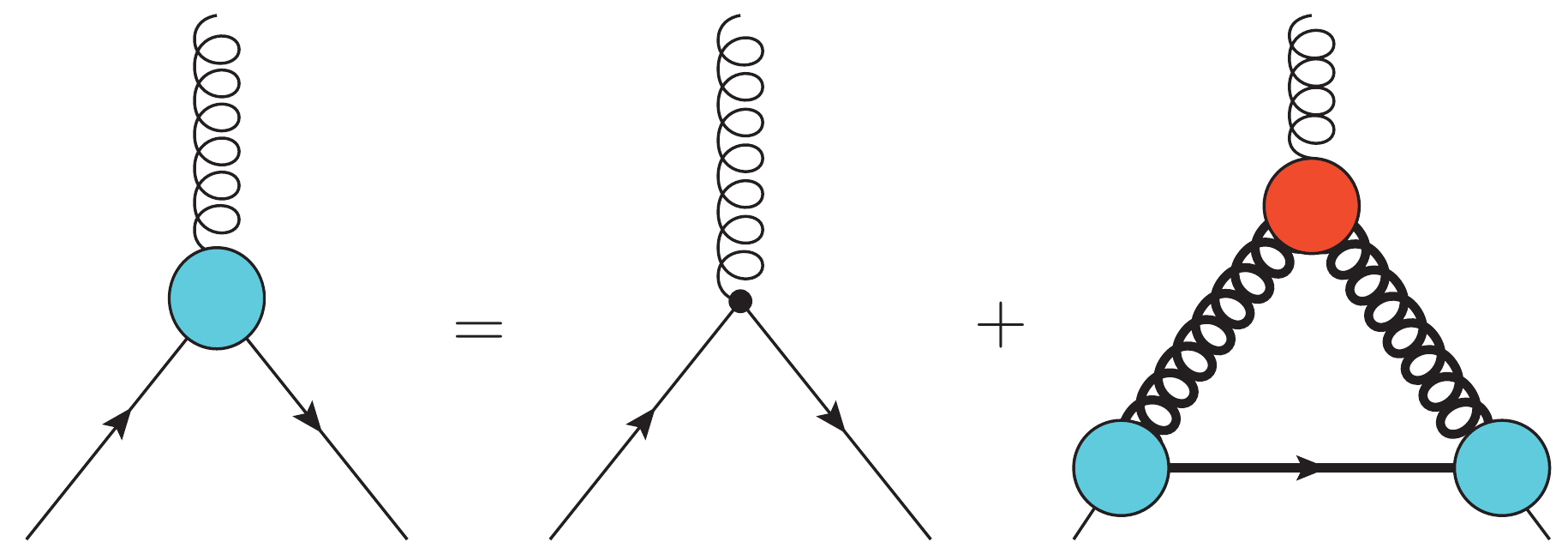}}
\caption{The truncated quark-gluon vertex DSE including the full non-Abelian diagram. 
For the three-gluon vertex model a modified version compared to 
the one used in \cite{Alkofer:2008tt,Windisch:2012de} has been employed.}
\label{Fig3}
\end{figure}

Since previous investigations of the system indicate a possible dominance of the non-Abelian 
diagram, the following simplified system depicted in Figs.\ \ref{Fig2} and \ref{Fig3} has 
been considered as a first step towards the full solution of the more 
complicated system given in Fig. \ref{Fig1}.
Using an even simpler approximation, a similar system has been treated previously \cite{Windisch:2012de}.
But in addition to \cite{Windisch:2012de} now the full non-Abelian
diagram has been included. In Fig. \ref{Fig3}, the complete tensor structure 
is coupled back to the full vertex,
where this object also enters the quark propagator equation in Fig. \ref{Fig2}.
We employ twelve tensor structures to span the vertex, where it turned out to be numerically advantageous 
to use a simple basis given by
\begin{equation}
\Gamma^\mu(p,q)\propto\left\{\begin{array}{c}
\mathbbm{1}\\
\slashed p\\
\slashed q\\
\frac{1}{2}{[\slashed p,\slashed q ]}
\end{array}\right\}\otimes\left\{\begin{array}{c}
\gamma^\mu\\
p^\mu\\
q^\mu
\end{array}\right\}\nonumber
\end{equation}
instead of the widely-used Ball-Chiu basis \cite{Ball:1980ay}.\footnote{A more efficient way to arrange
the basis elements can be found in \cite{Eichmann:2012mp}. This choice of basis setting will be employed in 
upcoming studies. We thank Richard Williams and Gernot Eichmann for discussions and 
useful hints.} Here, $p$ and $q$ are the in- and outgoing quark momenta.
As input from the Yang-Mills sector a gluon propagator has been employed 
which is quenched and of scaling type \cite{Alkofer:2008tt,Fischer:2002hna,Hopfer:2012ht}.\footnote{It can be
shown that unquenching effects play a minor role at vanishing temperatures \cite{Fischer:2003rp}. Nevertheless,
studying the back-reaction of quarks on the Yang-Mills sector by replacing the 
usual Ball-Chiu- or Curtis-Pennington vertex constructions with a full self-consistent 
quark-gluon vertex would be an
interesting task. In particular when going to finite chemical potential and temperature in
future calculations this step will become mandatory. Furthermore, we want to stress again that a 
gluon propagator of decoupling type could be employed here as well since the deep infrared behavior of the 
Yang-Mills system should not affect the phenomenological results.} 
Here we note that instead of using fits \cite{Alkofer:2008tt} for the gluon propagator, 
it is convenient to include this object also from a direct calculation \cite{Hopfer:2012ht} since this step
leaves the possibility to take more recent findings for the Yang-Mills system into account, 
see {\it e.g.}, 
\cite{Huber:2012kd,Cucchieri:2008qm,Alkofer:2004it,Alkofer:2008dt} and references therein.

For the three-gluon vertex in Fig. \ref{Fig3} a similar approach as in \cite{Huber:2012kd} has been employed.
It is given by the tree-level tensor structure of this object combined with the following bose-symmetric ansatz
\begin{equation}
\Gamma^{3g}(x)=-\left(\frac{x}{d_1+x}\right)^{-3\kappa}\left(\frac{d_1}{d_1+x}\right)^4 + 
d_2\,\log\biggl[\frac{x}{2}+1\biggr]^{17/44}
\end{equation}
where $x=p_1^2+p_2^2+p_3^2$ is the sum over the squared momenta entering the vertex and 
$\kappa \approx 0.595$.
This ansatz inherits the strong infrared enhancement known 
from Yang-Mills theory (see however the footnote in Sec.\ref{sec:intro})
as well as ensures the correct anomalous dimension of the vertex in the UV region. The novel improvement
compared to \cite{Alkofer:2008tt,Windisch:2012de} is the zero crossing for small momenta
such that  $\Gamma^{3g}$ becomes negative for small momenta. 
This behavior is strongly indicated by recent lattice calculations performed in two, three 
and four space-time dimensions, see \cite{Huber:2012kd} for a detailed discussion. 
In two space-time dimensions, the results
show a zero crossing such that the three-gluon 
vertex becomes negative \cite{Maas:2007uv}
and are furthermore in agreement with the expected scaling behavior \cite{Huber:2012zj}, whereas
in three and four dimensions 
no scaling solution has been found on the lattice so far. Thus, the three-gluon vertex might 
approach a negative but finite value in the infrared regime in these cases as pointed out in \cite{Huber:2012kd}. 
Unfortunately in four 
space-time dimensions the statistics is not yet high enough to draw definite conclusions about the 
deep infrared region and therefore future lattice calculations are needed to clarify this issue.
Nevertheless, the vertex is suppressed towards small momenta and shows qualitative 
agreement with the three-dimensional case where this suppression is
even more pronounced and a clear sign flip has been observed \cite{Cucchieri:2008qm}.\footnote{Furthermore, 
due to the qualitative similarity of the propagators in three and four space-time dimensions there
is evidence that the zero crossing might also occur in four space-time dimensions as pointed out in 
\cite{Huber:2012kd}.}
Therefore we included this property also in our model for the three-gluon vertex. In addition, 
we show in Sec. \ref{sec:results} that no stable solution could have been obtained without 
performing this step.
Additionally, the model inherits two residual parameters
$d_1$ and $d_2$ which have to be fixed to physical observables, e.g. the chiral condensate.
%%%%%%%%%%%%%%%%%%%%%%%%%%%%%%%%%%%%%%%%%%%%%%%%%%%%%%%%%%%%%%%%%%%%%%%%%%%%%%%%%%%%%%%%%%%%%%%%%%%%%%%%%%%%%%%%%

\subsection{Obtaining the vertex dressing functions}
Here, we follow the procedure as proposed in \cite{Windisch:2012de}.
As a first step, a disentanglement of the dressing functions is necessary in order to get them
in an explicit form. Since the employed basis is neither orthonormal nor orthogonal 
a linear system of equations for the vertex dressing functions 
has to be solved which can be calculated in advance yielding explicit
expressions for the dressing functions as linear combinations of the twelve right-hand side
projections. The algebraic manipulations have been performed with the program 
\texttt{FORM} \cite{Vermaseren:2000nd}. During each iteration step these projections are calculated
numerically, and subsequently, are plugged into the solution of the pre-calculated linear system
in order to obtain the vertex dressing functions.  

\subsection{Renormalization and numerical treatment}
The renormalization procedure is performed analogous to \cite{Windisch:2012de},
except that the renormalization constant $Z_{1F}$ in
front of the non-Abelian diagram is now absent due to the 3PI structure of the vertex equation.
The mass renormalization constant $Z_m$, the quark wave function renormalization constant $Z_2$
as well as the quark-gluon vertex renormalization constant $Z_{1F}$ in front of the bare vertex 
have been fixed within a momentum subtraction (MOM) scheme.
In particular, the renormalization constants are fixed by the conditions 
$A(\mu^2)=\lambda_1(\mu^2,\mu^2,2\mu^2)=1$, where $\mu^2$ denotes some large renormalization scale.
Note that only $\lambda_1$, corresponding to the tree-level part $\gamma^\nu$, has to be 
treated in the renormalization process since
all other tensor structures lead to UV finite integrals due to the power-like
suppression of the corresponding contributions at large momenta. 
Furthermore, the condition $B(\mu^2)=m_R$ fixes the constant $Z_m$, 
where $m_R$ is the mass at the renormalization scale.

The evaluation of the coupled system depicted in Fig. \ref{Fig2} and Fig. \ref{Fig3} is numerically expensive.
Thus, the actual calculations are performed on Graphics 
Processing Units (GPUs) using CUDA\texttrademark and openMPI. 
In fact, the parallelization of the problem is straightforward since a decomposition into 
independent smaller fractions is possible. Each vertex dressing function 
$\lambda_{i=1\ldots12}(p^2,q^2,p\cdot q)$ depends on three variables.
Therefore, we employ a coarse grained parallelization using openMPI
corresponding to the external momentum $p^2$ in the quark propagator. On each GPU a fine 
grained parallelization using CUDA\texttrademark has been employed which performs
the subsequent integrals for all twelve dressing functions and the particular value of $p^2$.
 After this step the synchronization 
and the evaluation of the linear system for the dressing functions takes place. The advantage of 
this procedure is the minimized communication between the individual processes
and the scaling to an, in principle, arbitrary number of GPUs.\footnote{In fact, the number of 
GPUs is limited only by the size of the external momentum grid, which ranges from 32 to 64 Gauss-Legendre
nodes in our calculations.}
%%%%%%%%%%%%%%%%%%%%%%%%%%%%%%%%%%%%%%%%%%%%%%%%%%%%%%%%%%%%%%%%%%%%%%%%%%%%%%%%%%%%%%%%%%%%%%%%%%%%%%%%%%%%%%%%%
\section{Results}
\label{sec:results}
\begin{figure}[h]
\centering
\subfigure{\includegraphics[width=7.05cm]{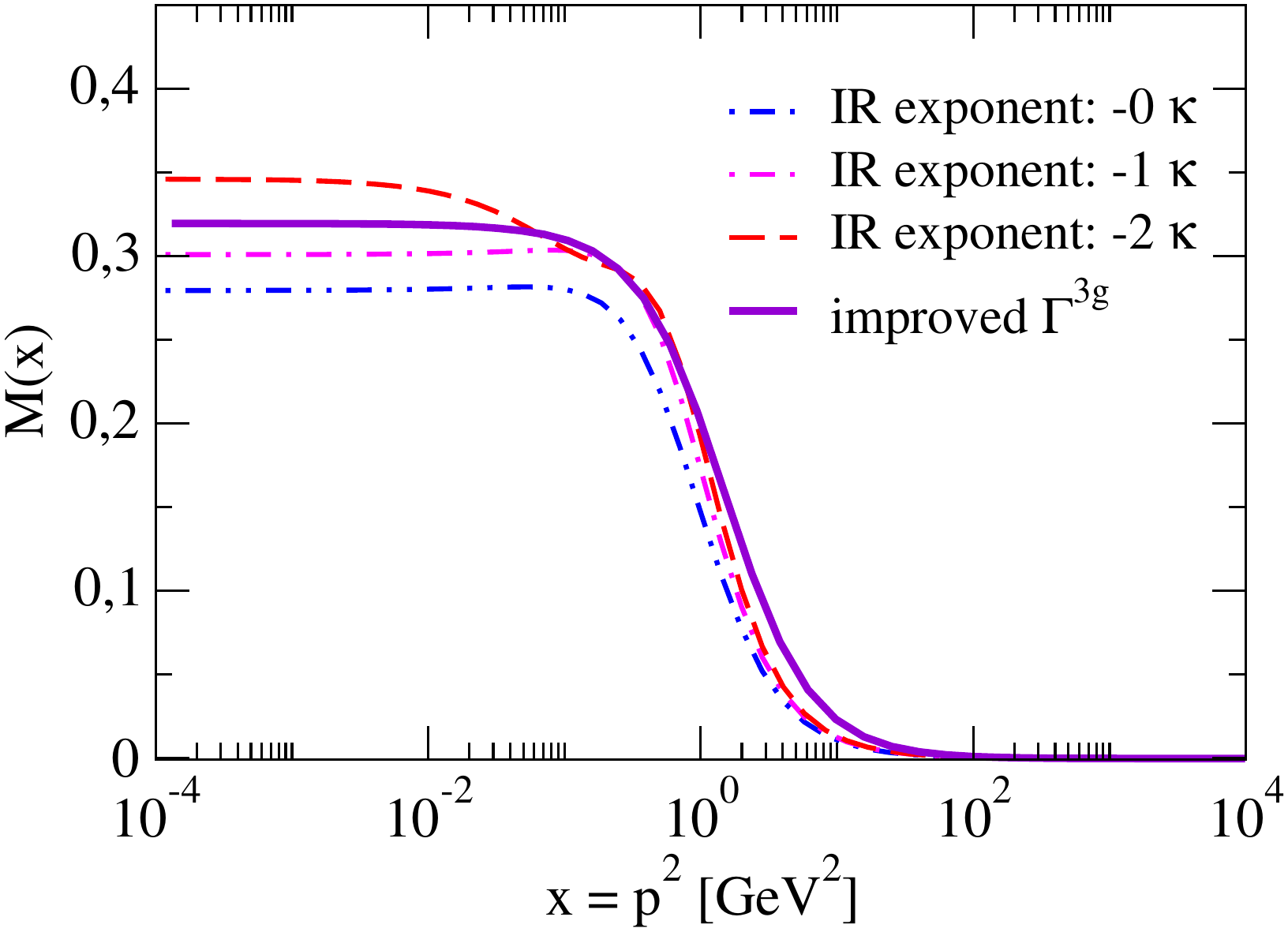}\label{fig:mass}}
\subfigure{\includegraphics[width=7.95cm]{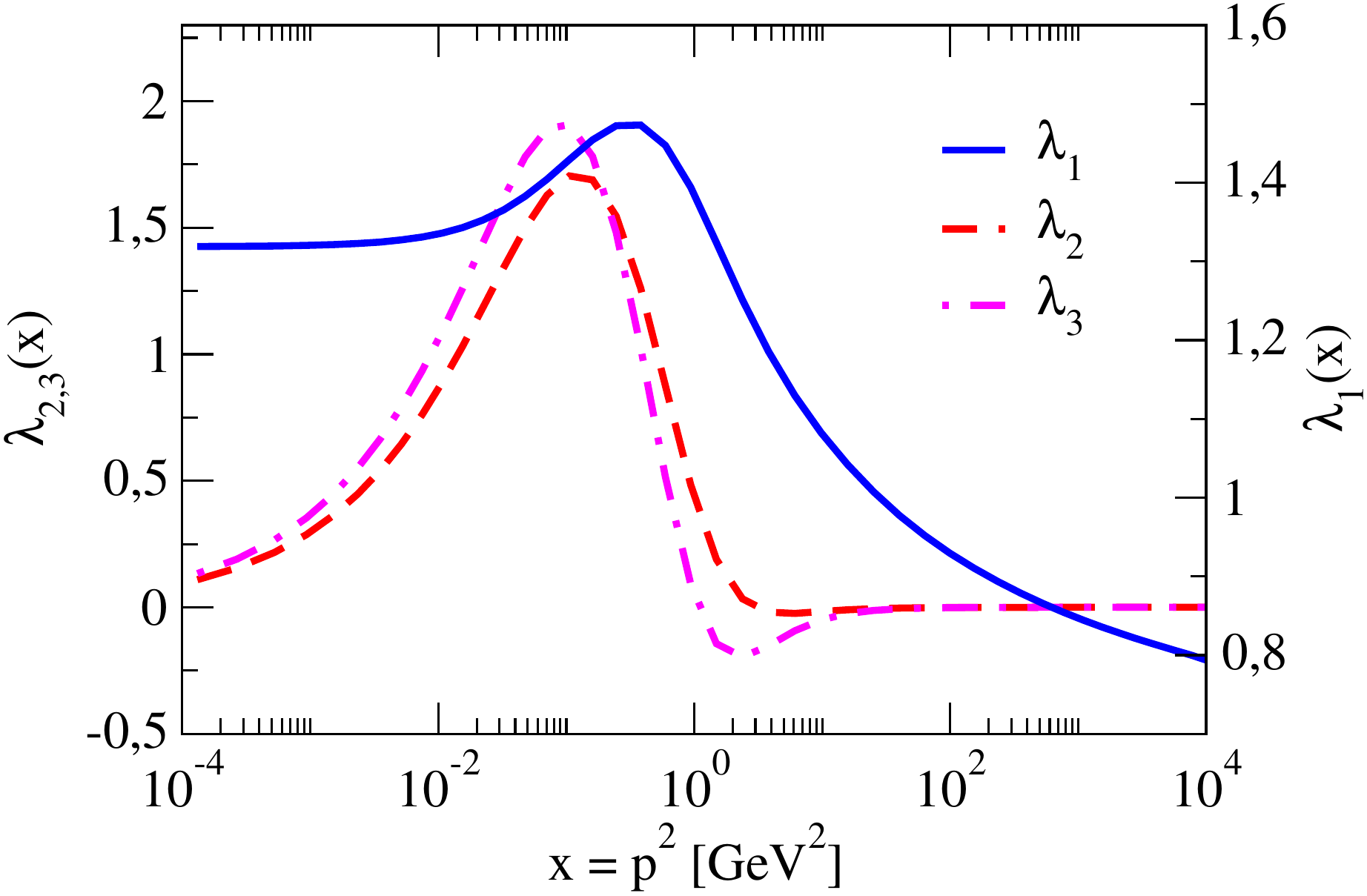}\label{fig:vertexdressings}}
\caption{
Left: The solid line shows the mass function $M(p^2)=B(p^2)/A(p^2)$ as obtained from the 
coupled system presented in Sec. 2.1. 
The dashed lines correspond to a previous treatment of the system \cite{Windisch:2012de} 
including only a partially dressed non-Abelian diagram. 
Here, the infrared exponent of the employed three-gluon vertex model \cite{Alkofer:2008tt,Windisch:2012de}
has been varied, where no stable solutions could have been obtained for values larger 
than $-2\kappa$. 
Right: The dimensionless vertex dressing functions $\lambda_1$, $\lambda_2$ and $\lambda_3$
corresponding to the tensor structures $\gamma^\nu$, $p^\nu$ and $-q^\nu$ of the 
quark-gluon vertex evaluated at $p^2=q^2$ and $p.q=0$.}
\end{figure}

The system described in Sec. \ref{sec:towards_a_solution} has been solved self-consistently,
taking all twelve tensor structures for the quark-gluon vertex into account. 
Due to stability issues it was in our previous study \cite{Windisch:2012de} not possible to take
the complete non-Abelian diagram as well as the full infrared strength of the three-gluon vertex
 $(p^2)^{-3\kappa}$ into account as can be seen in Fig. \ref{fig:mass} which shows the behavior of the
mass function $M(p^2)$ as obtained from the quark propagator dressing functions in the chiral limit.
Here, the corresponding infrared exponent of the three-gluon vertex has been varied from $0$ to $-2\kappa$ 
resulting in moderate changes of the dynamically generated mass (dashed lines). 
The solid line shows the dynamically generated mass as obtained from the full treatment of the system described 
in Sec. \ref{sec:towards_a_solution}.
Fig. \ref{fig:vertexdressings} shows the dimensionless vertex dressing functions $\lambda_1$, $\lambda_2$
and $\lambda_3$ 
which correspond to the
tensor structures $\gamma^\nu$, $p^\nu$ and $-q^\nu$ evaluated at the symmetric point $p^2=q^2$ and $p.q=0$.
We note that we observe the significant infrared enhancement
also for the other vertex tensor structures, independent whether they are invarariant under 
chiral transformations or are non-invariant and thus (in the chiral limit)
 generated dynamically via D$\chi$SB.
%\footnote
{Furthermore, the decoupling type of solution
seen in the calculations might be attributed to a different initialization strategy of the vertex 
dressing functions as compared to the one in ref.\  \cite{Alkofer:2008tt}.}
%%%%%%%%%%%%%%%%%%%%%%%%%%%%%%%%%%%%%%%%%%%%%%%%%%%%%%%%%%%%%%%%%%%%%%%%%%%%%%%%%%%%%%%%%%%%%%%%%%%%%%
\section{Conclusions and outlook}
The coupled system of quark propagator and quark-gluon vertex has been 
investigated in Landau gauge using a Dyson-Schwinger approach. Results for the quark propagator 
coupled the quark-gluon vertex DSE including the full non-Abelian diagram have been presented.
An enhancement in the infrared regime has been observed in most of the quark-gluon vertex dressing
functions leading to an effective quark-gluon interaction strength strong enough to generate mass dynamically.
The isolation of relevant tensor structures seems feasible and will be investigated in an upcoming
study of the system, where additionally the influence of the Abelian diagram within a dynamical setup
will be investigated, thus aiming 
towards a complete solution of the system as depicted in Fig.~\ref{Fig1}. 
Isolating important tensor structures and thus a reduction of complexity is also mandatory in studies involving 
the quark-gluon vertex, especially those at non-vanishing temperatures and/or quark chemical potentials.
%%%%%%%%%%%%%%%%%%%%%%%%%%%%%%%%%%%%%%%%%%%%%%%%%%%%%%%%%%%%%%%%%%%%%%%%%%%%%%%%%%%%%%%%%%%%%%%%%%%%%%
\section*{Acknowledgments}
We thank Gernot Eichmann, Christian Fischer, Markus Q. Huber, Manfred Liebmann, Felipe Llanes-Estrada, 
Mario Mitter and Richard Williams for helpful discussions. MH and AW acknowledge support 
by the Doktoratskolleg ''Hadrons in Vacuum, Nuclei and Stars`` of the Austrian Science Fund, 
FWF DK W1203-N16. This study is also supported by the 
Research Core Area ''Modeling and Simulation`` of the University of Graz, Austria.  
%%%%%%%%%%%%%%%%%%%%%%%%%%%%%%%%%%%%%%%%%%%%%%%%%%%%%%%%%%%%%%%%%%%%%%%%%%%%%%%%%%%%%%%%%%%%%%%%%%%%%%

\end{document}